\documentclass[conference]{IEEEtran}
\IEEEoverridecommandlockouts
\usepackage{cite}
\usepackage{amsmath,amssymb,amsfonts}
\usepackage{algorithmic}
\usepackage{graphicx}
\usepackage{textcomp}
\usepackage{xcolor}
\usepackage{multirow}
\usepackage{booktabs}

\def\BibTeX{{\rm B\kern-.05em{\sc i\kern-.025em b}\kern-.08em
    T\kern-.1667em\lower.7ex\hbox{E}\kern-.125emX}}
\begin{document}

\title{LiteShield: Hybrid Feature Selection-Driven Lightweight Intrusion Detection for Resource-Constrained IoT Networks
}

\author{\IEEEauthorblockN{1\textsuperscript{st} Dileepa Mabulage}
\IEEEauthorblockA{\textit{Department of Computing} \\
\textit{Informatics Institute of Technology}\\
Colombo, Sri Lanka \\
dileepa.20212025@iit.ac.lk}
\and
\IEEEauthorblockN{2\textsuperscript{nd} Banuka Athuraliya}
\IEEEauthorblockA{\textit{Department of Computing} \\
\textit{Informatics Institute of Technology }\\
Colombo, Sri Lanka \\
banu.a@iit.ac.lk}
}

\maketitle

\begin{abstract}
The rapid expansion of Internet of Things (IoT) deployments has enlarged the attack surface of modern digital infrastructure while exposing a key security mismatch: many intrusion detection systems (IDSs) remain too computationally expensive for constrained IoT environments. This paper presents \textit{LiteShield}, a lightweight machine learning-based IDS that combines hybrid feature selection with efficient classifiers to support accurate attack detection under limited computational budgets. The proposed framework uses the UNSW-NB15 dataset, applies data preprocessing and imbalance-aware preparation, and employs a two-stage feature selection pipeline based on Mutual Information (MI) and Recursive Feature Elimination with Cross-Validation (RFECV). Six lightweight classifiers are evaluated for both binary and multiclass intrusion detection: Decision Tree, Random Forest, K-Nearest Neighbors (KNN), Logistic Regression, Na\"ive Bayes, and Support Vector Machine.

Experimental results show that KNN achieved the highest raw predictive performance, reaching 98.26\% accuracy for binary classification and 85.22\% accuracy for multiclass classification. However, Random Forest delivered the most practical trade-off between detection quality and deployment efficiency, obtaining 98.01\% binary accuracy and 80.39\% multiclass accuracy with substantially lower model size and inference cost than KNN. Additional ablation analysis on minority attack classes indicates that class imbalance materially affects multiclass performance. Overall, LiteShield demonstrates that hybrid feature selection and lightweight machine learning can provide a viable path toward accurate and computationally feasible intrusion detection for IoT-focused environments.
\end{abstract}


\section{Introduction}
The Internet of Things has transformed domains such as healthcare, smart cities, transportation, and industrial automation, but the resulting ecosystems have also become highly exposed to cyberattacks \cite{agbedanu2022,nguyen2022}. IoT devices are typically constrained by limited memory, compute, bandwidth, and power, which makes deployment of conventional security controls difficult \cite{agbedanu2022,roy2022}. At the same time, modern intrusion detection increasingly relies on machine learning (ML) and deep learning (DL), yet many high-performing models remain too computationally intensive for realistic edge deployment \cite{benaddi2025,ahsan2022}.

This creates a design tension: an IoT IDS must preserve detection quality while remaining lightweight enough for constrained devices. Existing lightweight IDS studies report encouraging results, but several gaps remain. First, many approaches rely on single-stage filter methods or standard dimensionality reduction rather than hybrid feature selection \cite{benaddi2025,roy2022}. Second, several systems emphasize binary classification or a narrow subset of attack types, limiting practical utility in multiclass scenarios \cite{khanday2023,ahsan2022}. Third, dataset imbalance continues to impair minority-class detection \cite{basati2022,mukhaini2024}.

To address these issues, this paper proposes \textit{LiteShield}, a lightweight IDS built around three ideas: (i) a hybrid MI+RFECV feature selection pipeline, (ii) comparative evaluation of six lightweight ML classifiers, and (iii) explicit analysis of the trade-off between predictive performance and resource efficiency. The main contributions are as follows:
\begin{enumerate}
    \item A hybrid feature selection pipeline for reducing high-dimensional network traffic data while preserving discriminative information.
    \item A comparative binary and multiclass evaluation of lightweight ML models for IoT intrusion detection.
    \item A practical analysis showing that the best raw classifier is not always the best deployment candidate under resource constraints.
\end{enumerate}

\section{Related Work and Research Gap}

Recent studies show growing interest in designing intrusion detection systems that are both accurate and lightweight enough for IoT environments. A major direction in the literature is the use of classical machine learning models with dimensionality reduction or feature selection to reduce computational cost. For example, \cite{agbedanu2022} proposed an IDS that combines Incremental Principal Component Analysis (IPCA) with SAM-KNN, showing that careful reduction of the feature space can support deployment on constrained platforms while preserving strong detection capability. Similarly, \cite{roy2022} introduced a lightweight supervised IDS that integrates PCA, multicollinearity removal, and data sampling to improve efficiency and reduce false alarms. These studies demonstrate that feature reduction is central to lightweight IDS design, particularly when memory footprint and inference latency are critical.

Another prominent direction is the use of optimized ensemble or hybrid learning pipelines. In this area, \cite{zhang2025} developed a two-stage feature selection framework that combines Spearman Correlation Coefficient with Salp Swarm Optimization and then applies LightGBM for classification. This work highlights the benefit of combining a fast filter stage with a more selective optimization stage to improve predictive performance. Likewise, \cite{albulayhi2022} showed that combining multiple feature ranking strategies can improve the quality of selected features compared to relying on a single statistical criterion. These studies suggest that hybrid feature selection is promising because it can better balance computational efficiency and predictive power than purely filter-based methods.

Deep learning has also been explored for IoT intrusion detection, especially to improve representation learning for complex traffic patterns. For instance, \cite{benaddi2025} presented a lightweight CNN-BiLSTM framework with Chi-Square-based feature selection on the UNSW-NB15 dataset, while \cite{ahsan2022} evaluated deep models such as CNN, CNN-LSTM, and DenseNet for intrusion detection. These approaches demonstrate strong pattern learning capability, especially for nonlinear attack behaviors. However, despite their detection potential, deep architectures typically require higher memory, longer training time, and greater inference cost. Such characteristics may reduce their practicality for resource-constrained IoT nodes, gateways, or edge devices where lightweight deployment is a core requirement.

A further issue identified in the literature is that many studies emphasize binary classification or only a narrow subset of attacks. For example, \cite{khanday2023} focused primarily on DDoS-oriented detection, which is valuable for specific threat scenarios but does not fully address the broader requirement of distinguishing among multiple heterogeneous attack categories. In real IoT networks, intrusion detection systems must often identify not only whether traffic is malicious, but also what type of malicious behavior is present. This makes multiclass classification more challenging and more practically relevant than binary-only detection. However, the literature consistently reports that multiclass intrusion detection suffers when attack classes are highly imbalanced or when minority classes have only limited training examples \cite{mukhaini2024}.

These observations reveal several research gaps. First, although feature selection is widely used, many studies rely on single filter-based methods such as PCA, Chi-Square, or correlation analysis. While these methods are computationally efficient, they may overlook feature dependencies and interactions that wrapper-based methods can capture \cite{albulayhi2022,kaushik2022}. Second, although deep models may improve classification performance, they often introduce computational overhead that conflicts with the lightweight requirements of IoT deployment \cite{ahsan2022,benaddi2025}. Third, multiclass intrusion detection remains insufficiently addressed, especially for rare attack categories, where class imbalance can significantly reduce recall and increase misclassification \cite{khanday2023,mukhaini2024}. Finally, many prior studies prioritize predictive accuracy but provide limited discussion of model size, inference efficiency, and deployment suitability, which are essential in practical IoT security systems.

Motivated by these gaps, LiteShield focuses on three design objectives: (1) integrating hybrid feature selection to combine the speed of filter methods with the subset refinement capability of wrapper methods, (2) supporting both binary and multiclass intrusion detection, and (3) benchmarking detection performance alongside efficiency-related factors relevant to resource-constrained IoT settings. In this way, the proposed system aims to offer a more balanced solution between detection effectiveness and deployment feasibility.

\section{Methodology}
\subsection{Dataset and Preprocessing}
LiteShield uses the UNSW-NB15 dataset, a widely adopted benchmark containing modern normal and attack traffic with nine attack classes \cite{moustafa2015}. Following the dataset configuration used in prior work, the study uses the official train--test split and operates on 45 predictive features after removal of irrelevant identifiers and timestamps \cite{moustafa2015,pansari2024}. Data preprocessing includes cleaning, duplicate handling, categorical encoding, and feature scaling. For multiclass detection, a balanced training set is used to reduce the effect of severe class imbalance.

\subsection{Hybrid Feature Selection}
The proposed framework combines a filter stage and a wrapper stage. Mutual Information (MI) is first applied to rank features by dependency with the target labels. Recursive Feature Elimination with Cross-Validation (RFECV) then refines the subset by iteratively removing less useful features according to classifier performance. This two-stage design aims to preserve the speed advantages of filter methods while improving subset quality through wrapper-based optimization \cite{mustaqim2021,albulayhi2022}. The final model uses 20 selected features for binary classification and 20 selected features for multiclass classification.

\subsection{Classification and Evaluation}
Six lightweight classifiers are trained and evaluated: Decision Tree (DT), Random Forest (RF), KNN, Logistic Regression (LR), Na\"ive Bayes (NB), and Support Vector Machine (SVM). Binary classification distinguishes benign from malicious traffic. Multiclass classification predicts one of the attack categories or normal traffic. Performance is measured using accuracy, precision, recall, F1-score, and false positive rate (FPR), while deployment practicality is assessed using model size, inference latency, and memory usage \cite{tama2021,mallidi2025}.

\section{Results and Discussion}
Table~\ref{tab:results} summarizes the main quantitative findings. KNN achieved the highest predictive performance in both binary and multiclass settings. In binary detection, it obtained 98.26\% accuracy and an F1-score of 98.42\%. In multiclass detection, it reached 85.22\% accuracy and an F1-score of 87.44\%. However, this performance came with substantial cost: 62.92 MB model size for binary, 124.97 MB for multiclass, and the highest latency among all compared methods.

By contrast, Random Forest produced a stronger balance between effectiveness and efficiency. It delivered 98.01\% binary accuracy with a 1.90 MB model and 80.39\% multiclass accuracy with a 4.32 MB model. This makes RF far more attractive for practical IoT deployment despite KNN's superior raw accuracy. Decision Tree also remained lightweight and fast, but its multiclass performance was lower than RF and KNN.

\begin{table}[t]
\caption{LiteShield classification performance and deployment cost}
\label{tab:results}
\centering
\scriptsize
\setlength{\tabcolsep}{3pt}
\begin{tabular}{lcccccc}
\toprule
\multirow{2}{*}{Model} & \multicolumn{3}{c}{Binary} & \multicolumn{3}{c}{Multiclass} \\
\cmidrule(lr){2-4}\cmidrule(lr){5-7}
 & Acc. & F1 & Size (MB) & Acc. & F1 & Size (MB) \\
\midrule
DT  & 96.91 & 97.23 & 0.64 & 78.47 & 81.65 & 2.29 \\
RF  & 98.01 & 98.21 & 1.91 & 80.39 & 83.55 & 4.32 \\
KNN & 98.26 & 98.42 & 62.92 & 85.22 & 87.44 & 124.97 \\
SVM & 80.44 & 83.72 & 0.0013 & 59.53 & 66.18 & 0.0027 \\
LR  & 81.22 & 84.18 & 0.0014 & 59.48 & 66.16 & 0.0029 \\
NB  & 69.50 & 65.19 & 0.0018 & 43.26 & 50.50 & 0.0043 \\
\bottomrule
\end{tabular}
\end{table}

The results also reinforce the value of the hybrid feature selection strategy. By narrowing the feature space before classification, LiteShield maintains strong predictive performance without requiring heavyweight DL architectures. This is particularly important because computational feasibility matters as much as raw accuracy in edge settings \cite{nguyen2022,misrak2025}. The results additionally align with broader literature showing that ensemble and tree-based methods often provide robust performance for network intrusion detection \cite{mahmud2024,zhang2025}.

\subsection{Ablation and Validation}
Ablation analysis on minority classes (Worms and Shellcode) further clarified the effect of class imbalance. When these classes were removed, accuracy increased across all models, indicating that underrepresented attacks degrade multiclass learning stability. For example, RF improved from 80.39\% multiclass accuracy in the full setting to 80.62\% when both Worms and Shellcode were removed, while KNN improved from 85.22\% to 85.34\%. Although the gains are modest, their consistency supports the claim that imbalance handling remains essential for realistic multiclass IDS design.

\begin{table}[t]
\caption{Multiclass accuracy under attack-class ablation}
\label{tab:ablation}
\centering
\scriptsize
\setlength{\tabcolsep}{4pt}
\begin{tabular}{lcccc}
\toprule
Model & Full & -Worms & -Shellcode & -Both \\
\midrule
DT  & 78.47 & 77.30 & 79.53 & 78.46 \\
RF  & 80.39 & 80.48 & 80.85 & 80.62 \\
KNN & 85.22 & 85.23 & 85.34 & 85.34 \\
LR  & 59.48 & 60.80 & 62.66 & 63.95 \\
NB  & 43.26 & 43.59 & 43.42 & 45.55 \\
SVM & 59.53 & 59.43 & 59.50 & 60.23 \\
\bottomrule
\end{tabular}
\end{table}

\subsection{Positioning Against Prior Work}
LiteShield's results compare favorably with several published baselines on UNSW-NB15 and related intrusion benchmarks. The binary RF result of 98.01\% accuracy is higher than values reported by Pansari \textit{et al.} (95.03\%) and Pal \textit{et al.} (87.53\%) \cite{pansari2024,pal2025}. Likewise, binary KNN reaches 98.26\%, exceeding several earlier KNN-based results. In multiclass evaluation, LiteShield's KNN result of 85.22\% surpasses Pansari \textit{et al.}'s 81.77\% but remains below the 93.06\% reported by Jouhari \textit{et al.} for a different hybrid setting. These comparisons suggest that LiteShield is competitive, especially when practical efficiency is considered together with predictive performance.

\section{Limitations}
This study has several limitations. First, evaluation relies primarily on UNSW-NB15, which may not fully capture the heterogeneity of real IoT deployments or evolving attack behavior. Second, the system was not validated on physical IoT hardware, so real-world deployment feasibility remains partly inferred from model size and runtime characteristics rather than directly observed. Third, while oversampling and balanced training help, rare and zero-day attacks remain difficult to model reliably. Finally, only lightweight ML models were evaluated; future work could examine quantized or compressed DL alternatives that preserve a low deployment footprint.

\section{Conclusion and Future Work}
This paper presented LiteShield, a lightweight IoT intrusion detection framework that combines hybrid feature selection with efficient machine learning classifiers. The results show that high detection performance is achievable without resorting to computationally expensive DL models. KNN delivered the best predictive scores overall, but Random Forest provided the most practical trade-off for constrained deployment because it preserved strong accuracy while remaining far smaller and faster.

The study confirms three main findings: hybrid feature selection is effective for reducing dimensionality without undermining detection quality; multiclass IoT intrusion detection remains strongly affected by class imbalance; and efficiency-aware benchmarking is necessary because the highest-accuracy model may be unsuitable for real deployment. Future work should extend evaluation to additional datasets such as BoT-IoT, CICIDS2017, and ToN-IoT, test the framework on real IoT hardware, support streaming data, and explore adaptive, explainable, and compressed learning methods for evolving attack environments.

\end{document}